\documentclass[slac_one]{revtex4}
\usepackage{graphicx}
\usepackage{fancyhdr}
\begin{document}

\title{Search for particles decaying into $Z\gamma$ at D0} 

%

\author{Alexey V. Ferapontov (for the D0 collaboration)}
\affiliation{Kansas State University, Manhattan, Kansas 66506, USA}

\begin{abstract}
Results on the search for particles decaying into $Z\gamma (\rightarrow \ell\ell\gamma$) are presented. 
Using roughly 1 fb$^{-1}$ of data, dilepton-plus-photon invariant mass distributions have been examined for 
an excess over the theoretical predictions. Having observed a good agreement between data and the standard model 
prediction we set 95\% C.L. upper limits on the cross section times branching fraction ($\sigma \times {\cal B}$) 
of the resonance $Z\gamma$ production.
\end{abstract}

\maketitle

\thispagestyle{fancy}


\section{INTRODUCTION} 
\begin{figure}
\includegraphics[scale=0.21]{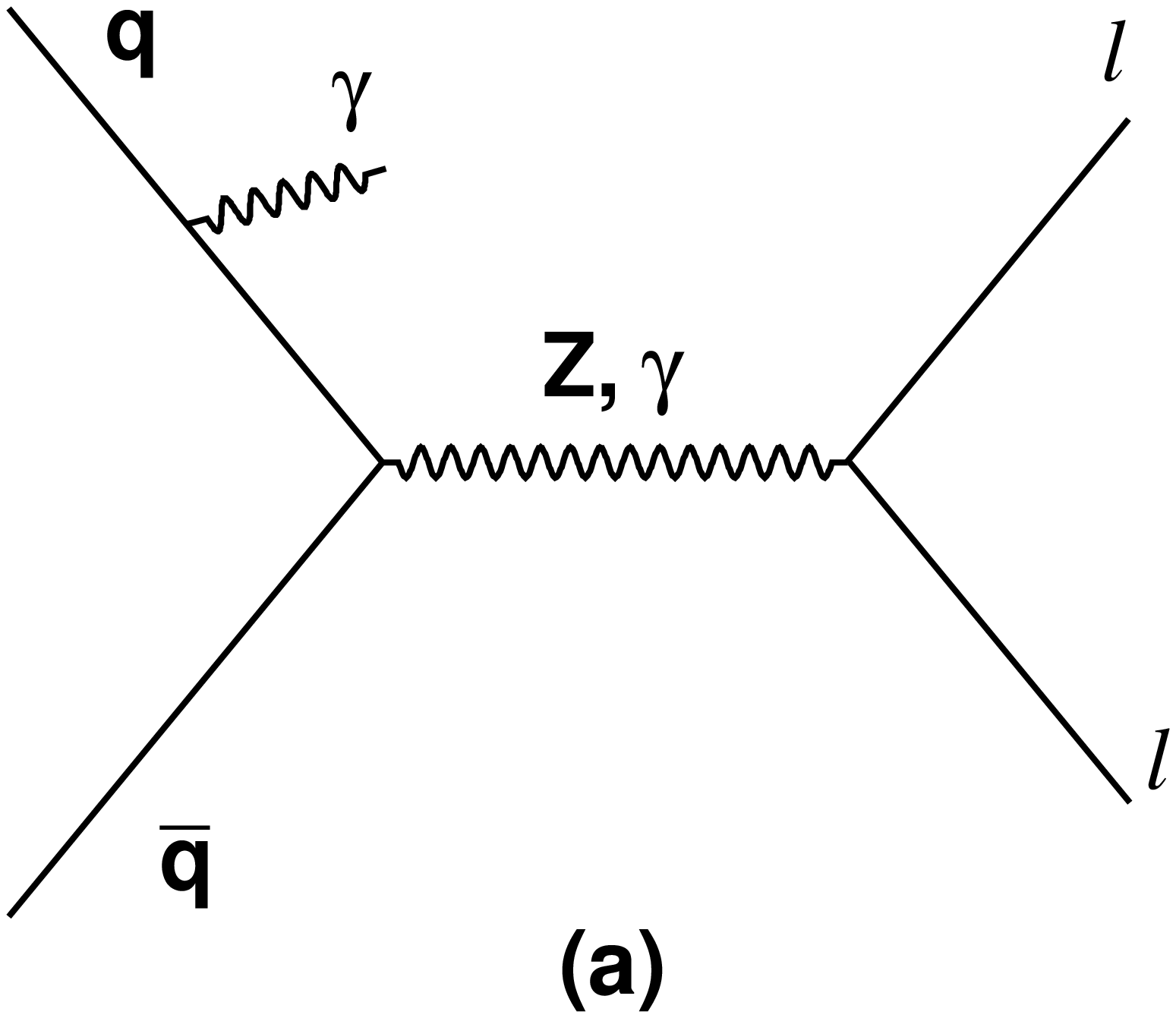} 
\includegraphics[scale=0.21]{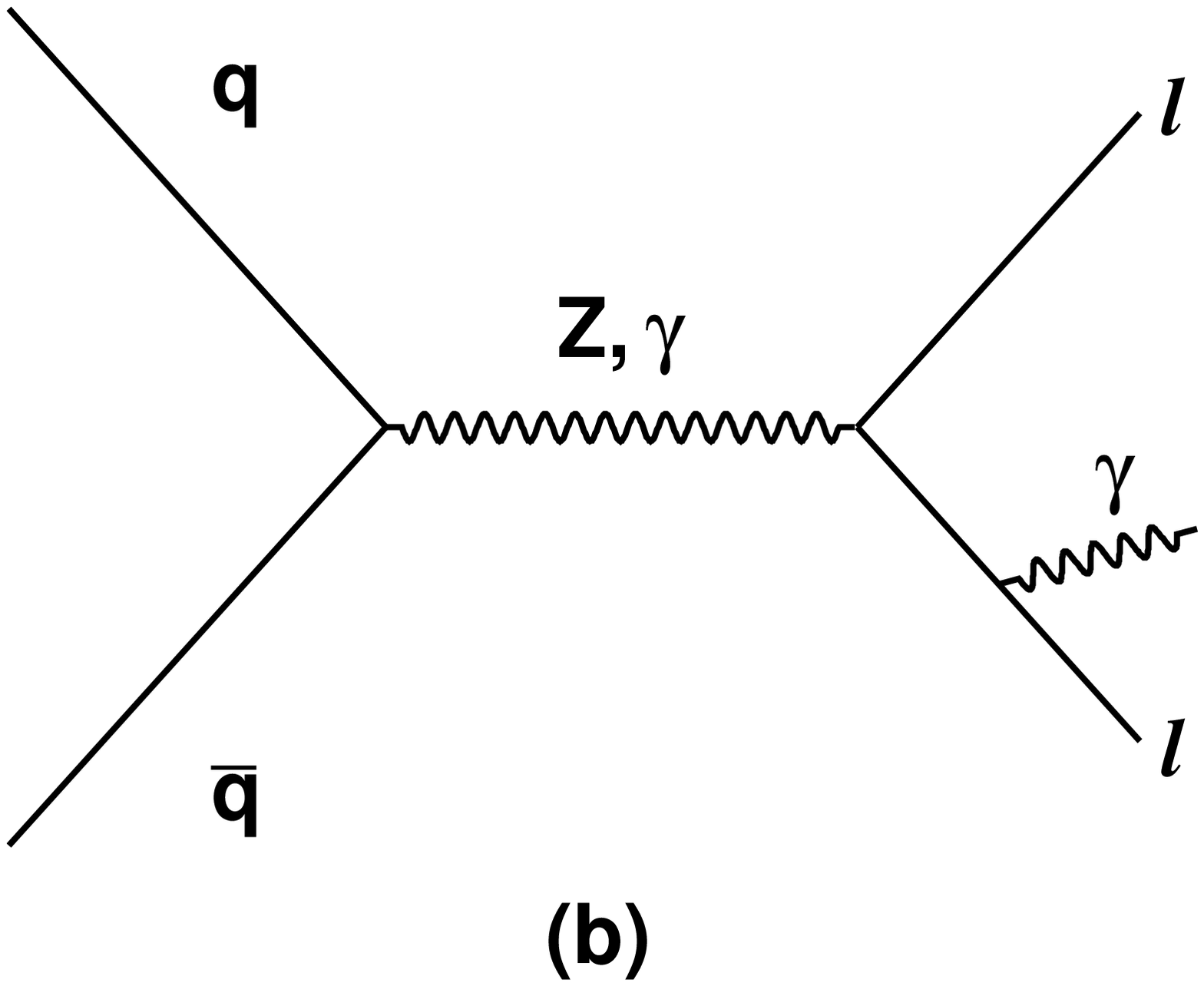} 
\includegraphics[scale=0.21]{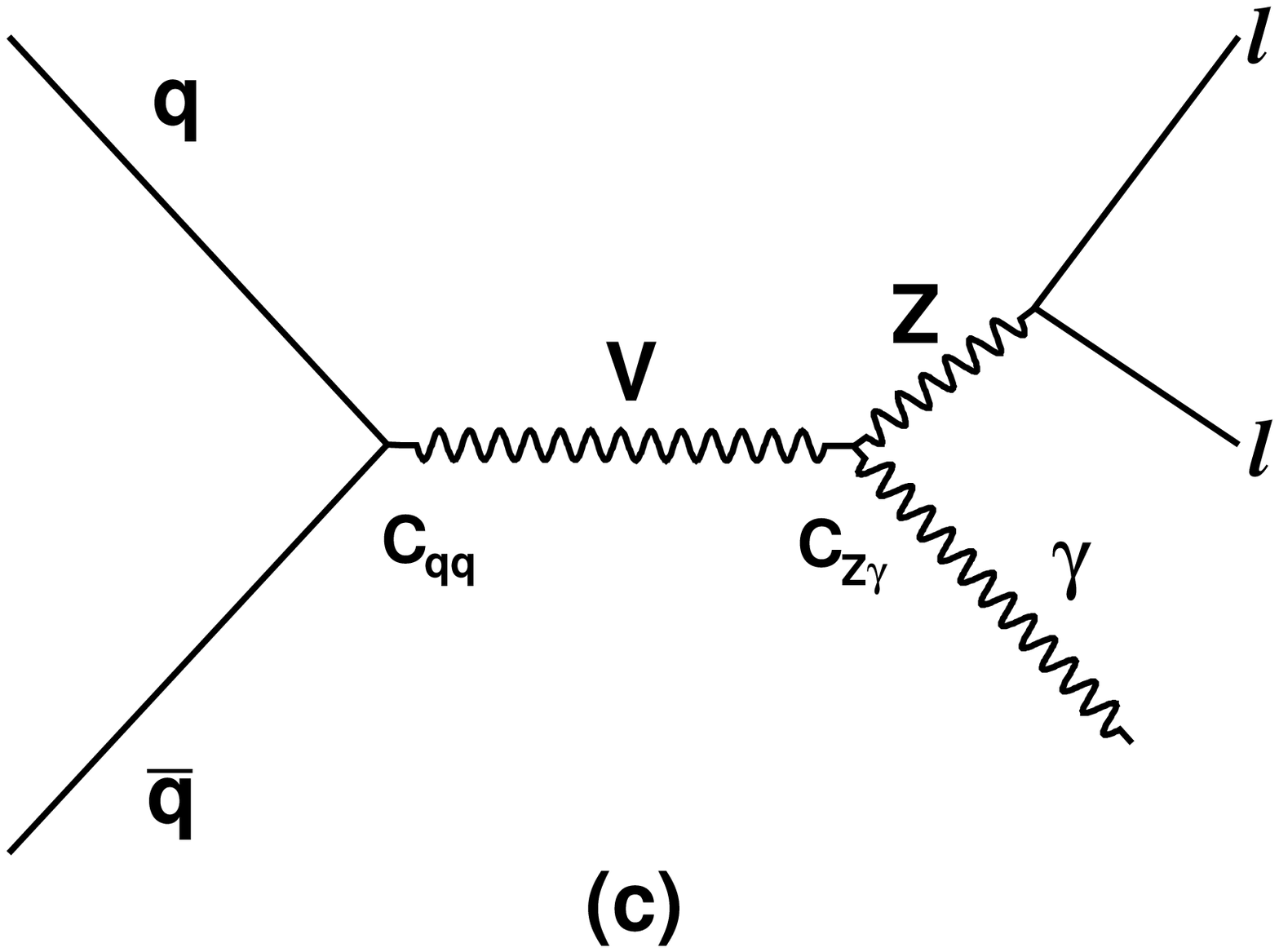} 
\includegraphics[scale=0.21]{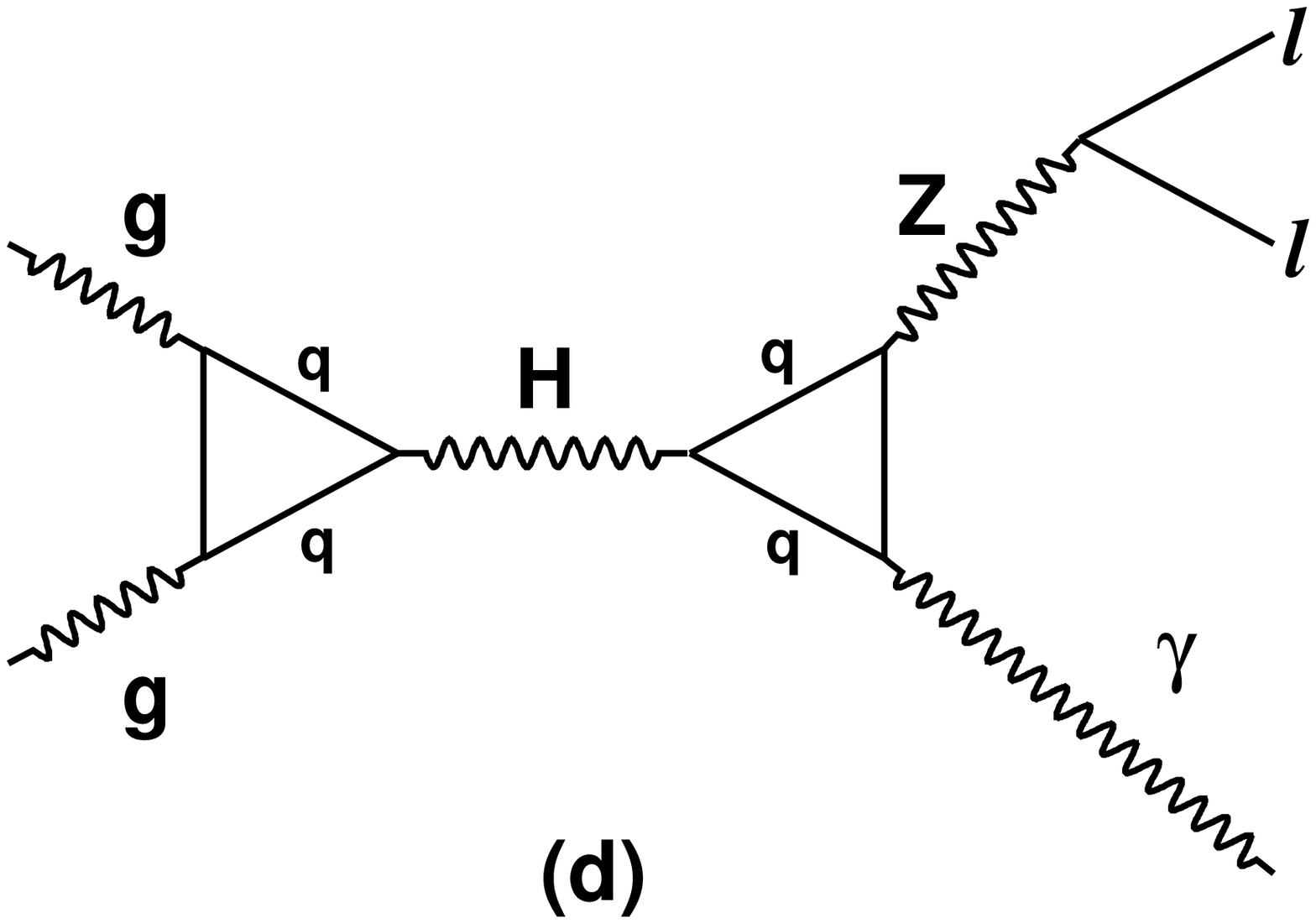}
\caption{Leading-order processes which produce $Z\gamma$ candidates:
(a) SM initial state radiation, (b) SM final state radiation, (c)
$q \bar q$ pair annihilation into a vector ($V$) particle which couples to the $Z \gamma$ and (d) SM Higgs production and decay.
\label{fig:feyn}}
\end{figure}

While the standard model (SM) describes the observed physics processes well at low and moderate energies, 
there is a possibility of existence of new physics beyond the SM, which manifests mostly at high energies. 
Studies of diboson production can lead to a potential discovery of $Z\gamma$ resonances predicted in a number of SM 
extensions: vector $Z^{\prime}$, scalar Higgs bosons, pseudo-scalar toponium, and techniparticles
~\cite{Buescher:2005re,Djouadi,Kozlov:2005rj,Ono:1983tf,Cakir:2004nh,Hill}. The D0 collaboration has already performed a search for such resonances in the following channels: $p\bar{p} \rightarrow X \rightarrow Z\gamma \rightarrow ee (\mu\mu) \gamma$ using 300 pb$^{-1}$ of data~\cite{us,Abazov:2006ez}. This current analysis~\cite{plb_subm} draws on the previous one, but only in the narrow resonance approximation: the total width of a resonance must be smaller than the detector resolution. The previous analysis is expanded by adding a vector resonance hypothesis. The scalar resonance is modeled with {\sc pythia}~\cite{pythia} using the SM Higgs boson production model, and {\sc madevent} \cite{madevent} is used to model a colorless, neutral, vector resonance. Diagrams for both processes are shown in Fig.~\ref{fig:feyn}. The D0 detector description can be found elsewhere~\cite{run2det}. 

\section{EVENT SELECTION}
The $Z\gamma$ final state is obtained by selecting events with a pair of energetic isolated leptons of the same 
family (electrons with transverse energy $E_T >$~15~GeV or muons with transverse momentum $p_T >$~15~GeV/$c$) and a photon candidate, separated from both leptons by $\Delta R = \sqrt{(\Delta \phi)^{2} + (\Delta \eta)^{2}} >$~0.7. Both electron and photon candidates are reconstructed from isolated~\cite{EM_isolation} 
electromagnetic (EM) showers in the calorimeter with at least 90\% of their energy deposited in the EM part of the calorimeter. In each event, photon candidate and at least one electron candidate are required to be identified in the central calorimeter. In addition, EM showers from electrons must be spatially matched to tracks found in the tracking system. By imposing an additional requirement on the sum of $p_T$ of all tracks, reconstructed in the annulus with opening angles of $R = $ 0.05 and 0.4 around the photon's trajectory, not to exceed 1.5 GeV/$c$, we substantially reduce the EM-like jets~\cite{emjets} background. In the muon channel, we select only those events in which muon candidates are reconstructed not back-to-back with an opening angle $\Delta\alpha_{\mu\mu} = |\Delta\phi_{\mu\mu} + \Delta\theta_{\mu\mu} - 2\pi| < 0.05$, and are produced in the same vertex. To further reduce backgrounds, we require muons to be isolated from activity in both calorimeter and tracking detectors. We also require the di-electron (dimuon) pair to be separated by $\Delta R_{ee} > 0.6$ ($\Delta R_{\mu \mu} > $ 0.5). To reduce systematic uncertainty due to the trigger efficiency, we require the $E_T$ of the leading electron candidate to exceed~25~GeV and the $p_T$ of the leading muon candidate to exceed 20 GeV/$c$ where the D0 triggers are fully efficient. The sensitivity to new physics resonances strongly depends on the detector resolution. Typical $M_{\ell\ell\gamma}$ resolution is on the order of 5\% in the electron channel, while in the muon channel this number varies from ~8\% to ~18\%, depending on the resonance mass. To improve the resolution of $M_{\mu\mu\gamma}$, we correct the transverse momenta of muon candidates that appear to be produced from the on-shell $Z$ boson by using the $Z$ boson nominal mass constraint. To further increase the analysis sensitivity, we vary cuts on photon transverse momentum and dilepton invariant mass to have the maximal $S/\sqrt{S+B}$ ratio. The resulting cuts are photon $E_T > $~20 GeV and dilepton mass $M_{\ell\ell}~>~$80~GeV/$c^2$.

\section{RESULTS}
Due to the fact that the total efficiency depends not only on the spin of the resonance (vector or scalar) 
but also on the lepton family (electron or muon), we study the following four cases separately: vector resonance 
in the electron or muon channel, and scalar resonance in the electron or muon channel. Afterwards, we combine 
electron and muon channel results for each type of resonance.
The total efficiency of the set of cuts described in the previous section multiplied by the 
geometrical and kinematic acceptance varies between 7\% and 20\% for both types of resonances.
These numbers include $\sim$60--68\% (79\%) electron (muon) identification efficiency per pair, 92--95\% 
photon reconstruction efficiency, and 99\% (68\%) electron (muon) trigger efficiency.

There are only two significant sources of background to the $Z\gamma$ final state: the SM $Z\gamma$
production and the $Z$+jet production, where an EM-like jet passes all photon selection 
criteria. The jet misreconstruction rate is an $E_T$-dependent function, and is defined 
as a ratio of EM-like jets that pass photon selection criteria to all EM objects that are reconstructed 
in the geometrical acceptance of the central calorimeter. The direct photon contribution, however, has to be 
subtracted from this rate. Furthermore, two data samples are used to estimate the $Z$+jet background: a sample that 
consists of real photons and EM-like jets, and a second one in which photon candidates fail the track isolation and 
photon shower shape requirements~\cite{us}. We estimate the $Z$+jet background to be $4.5 \pm 0.7({\rm stat.}) \pm 0.6({\rm syst.})$ events in the electron channel and $4.4 \pm 0.7({\rm stat.}) \pm 0.6({\rm syst.})$ events in the muon channel. The leading-order (LO) Baur event generator~\cite{baur} is used to estimate the number of SM $Z\gamma$ 
background events. The trilinear $Z\gamma\gamma$ and $ZZ\gamma$ couplings are set to their SM values (zero), and, 
in addition, the LO photon $E_T$~spectrum is corrected for the next-to-leading-order (NLO) effects using the 
$E_T$-dependent $K$-factor from the NLO Baur event generator~\cite{baur_NLO}. The SM $Z\gamma$ background is 
estimated to be $37.4 \pm 6.1({\rm stat.}) \pm 2.6({\rm syst.})$ events in the electron channel and $41.6 \pm 6.5({\rm stat.}) \pm 2.2({\rm syst.})$ events in the muon channel. Applying all the selection criteria in data yields 
49 $ee\gamma$ and 50 $\mu\mu\gamma$ candidate events.

\begin{figure}[t]
\includegraphics[scale=0.4]{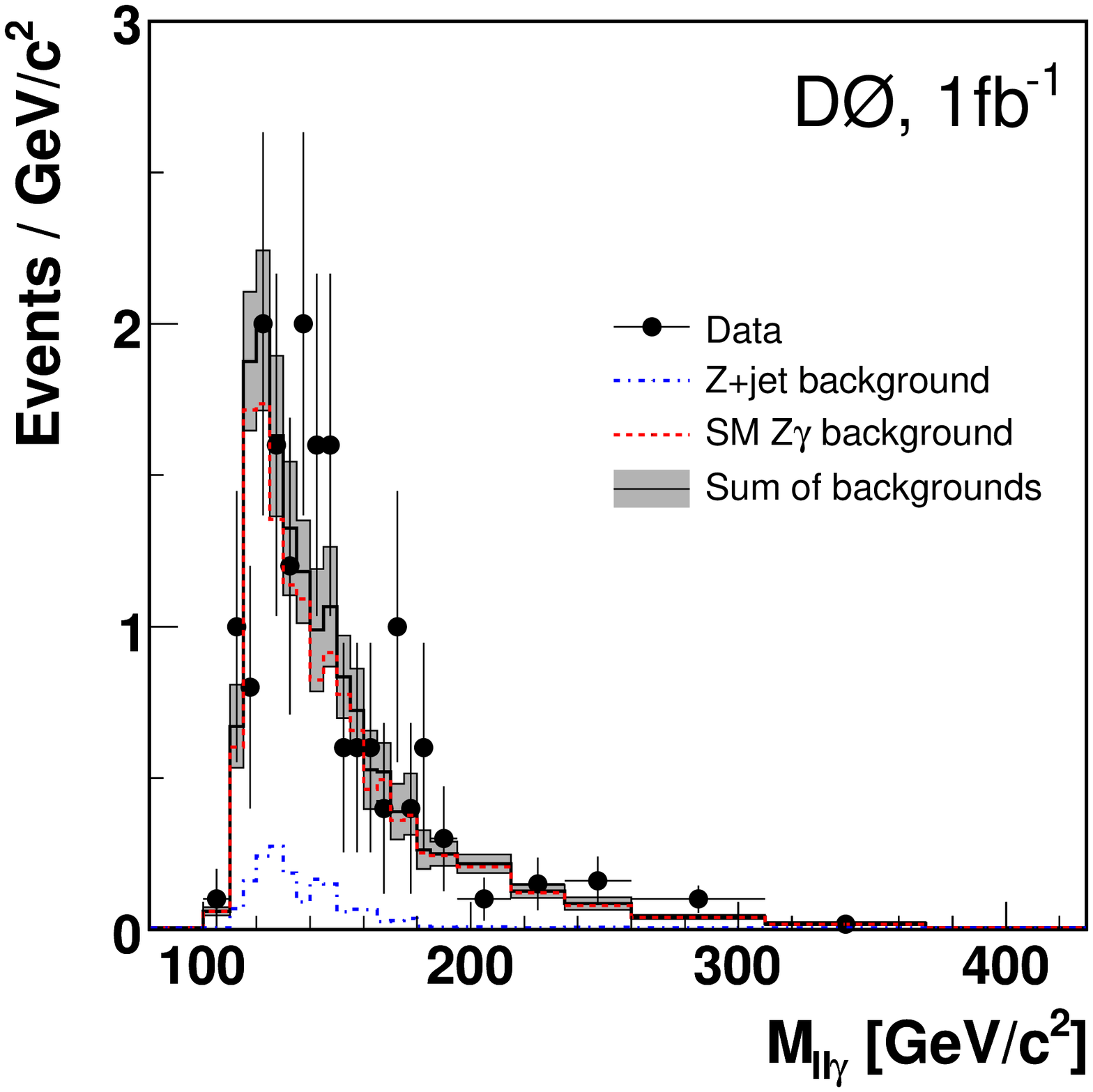} 
\includegraphics[scale=0.4]{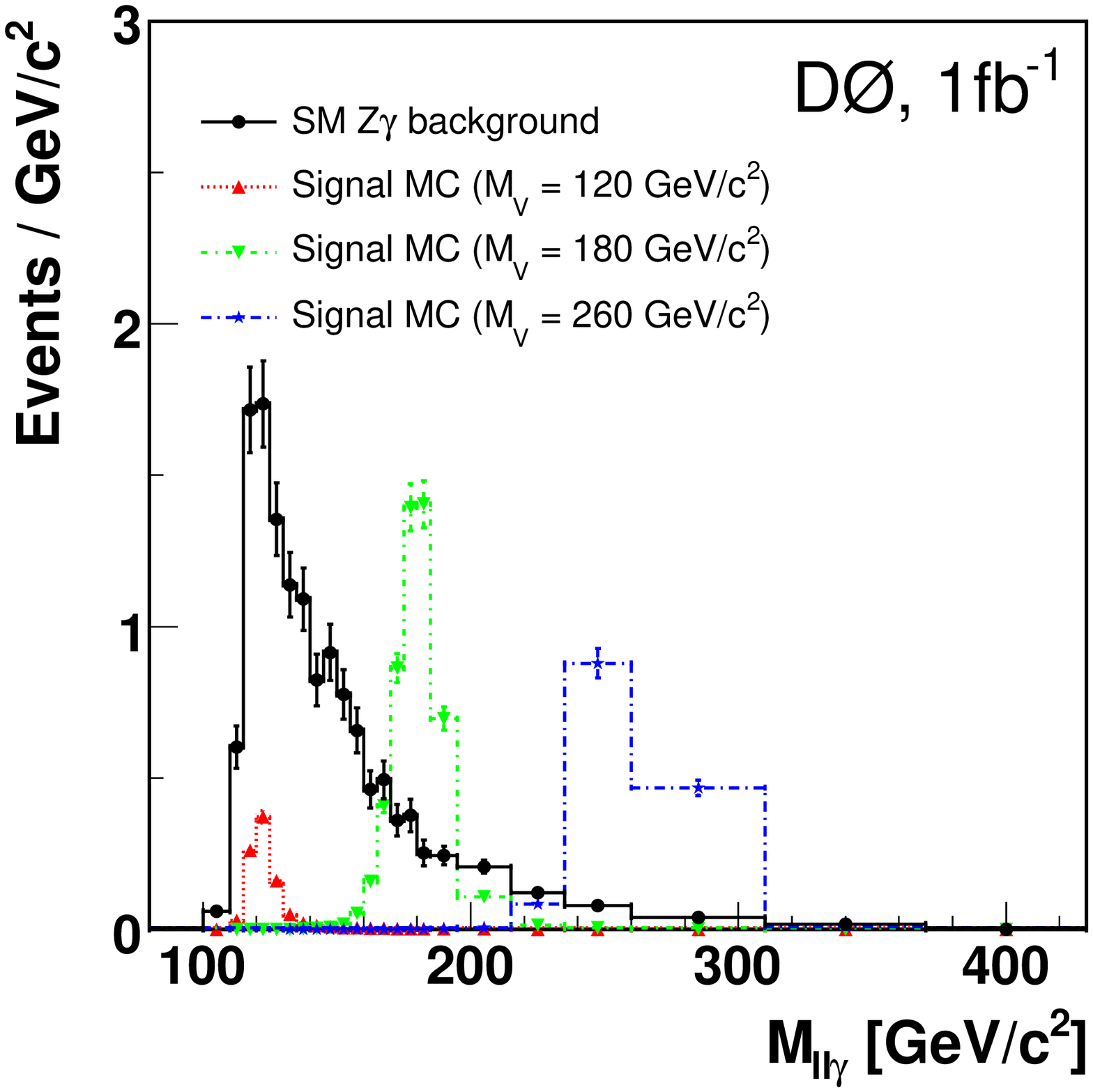}
\caption{
Left: invariant dilepton-plus-photon mass spectrum for $\ell\ell\gamma$ data (dots), 
SM $Z\gamma$ background (solid line histogram) and $Z$+jet background (dashed line histogram). The shaded band illustrates the systematic and statistical uncertainty on the sum of backgrounds. Right: MC signal $M_{\ell\ell\gamma}$ distributions of a vector particle for resonance masses of 120, 180 and 260~GeV/$c^2$.
\label{fig:mllg_vector}}
\end{figure}

\begin{figure}[t]
\includegraphics[scale=0.42]{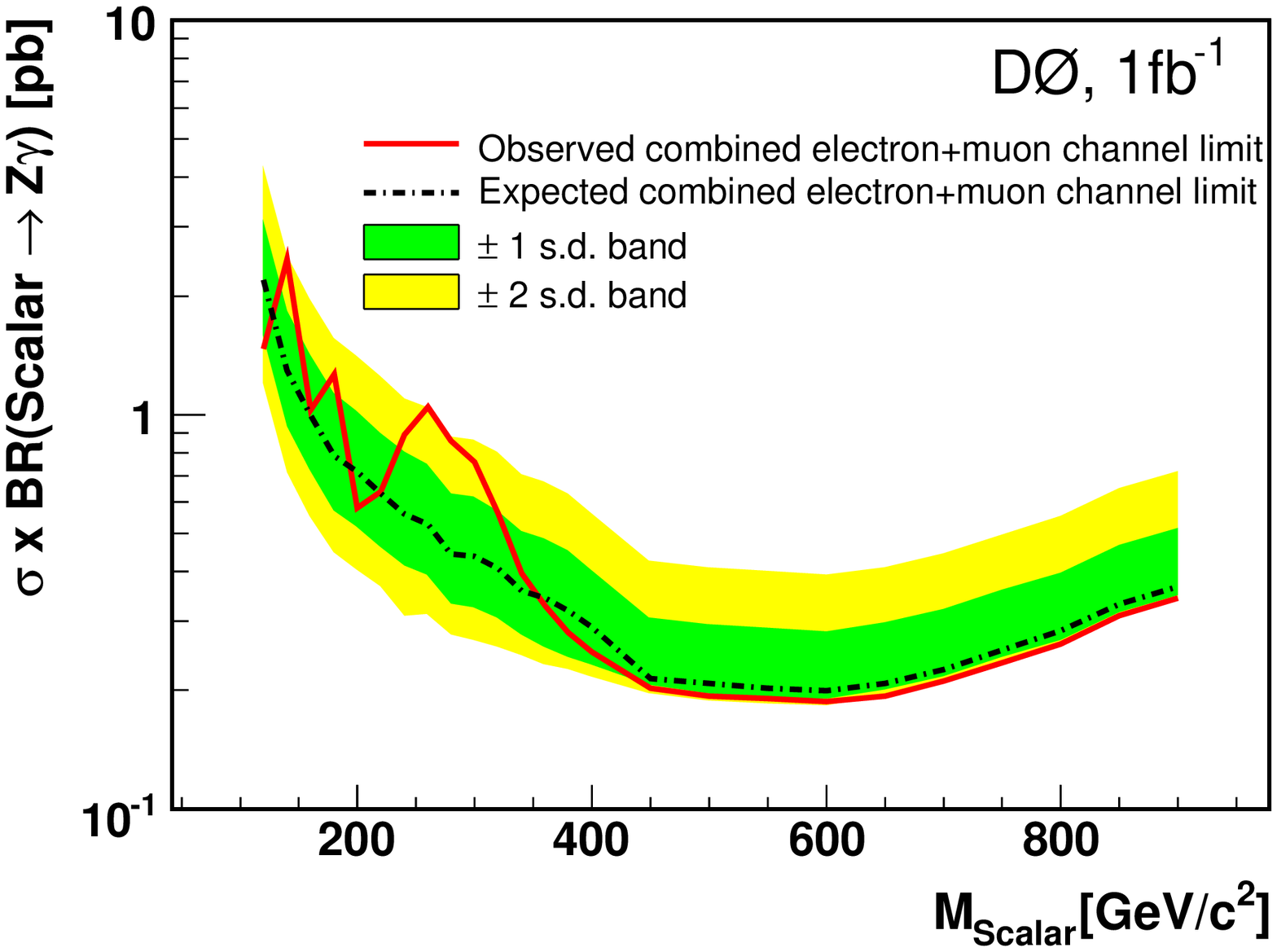}
\includegraphics[scale=0.42]{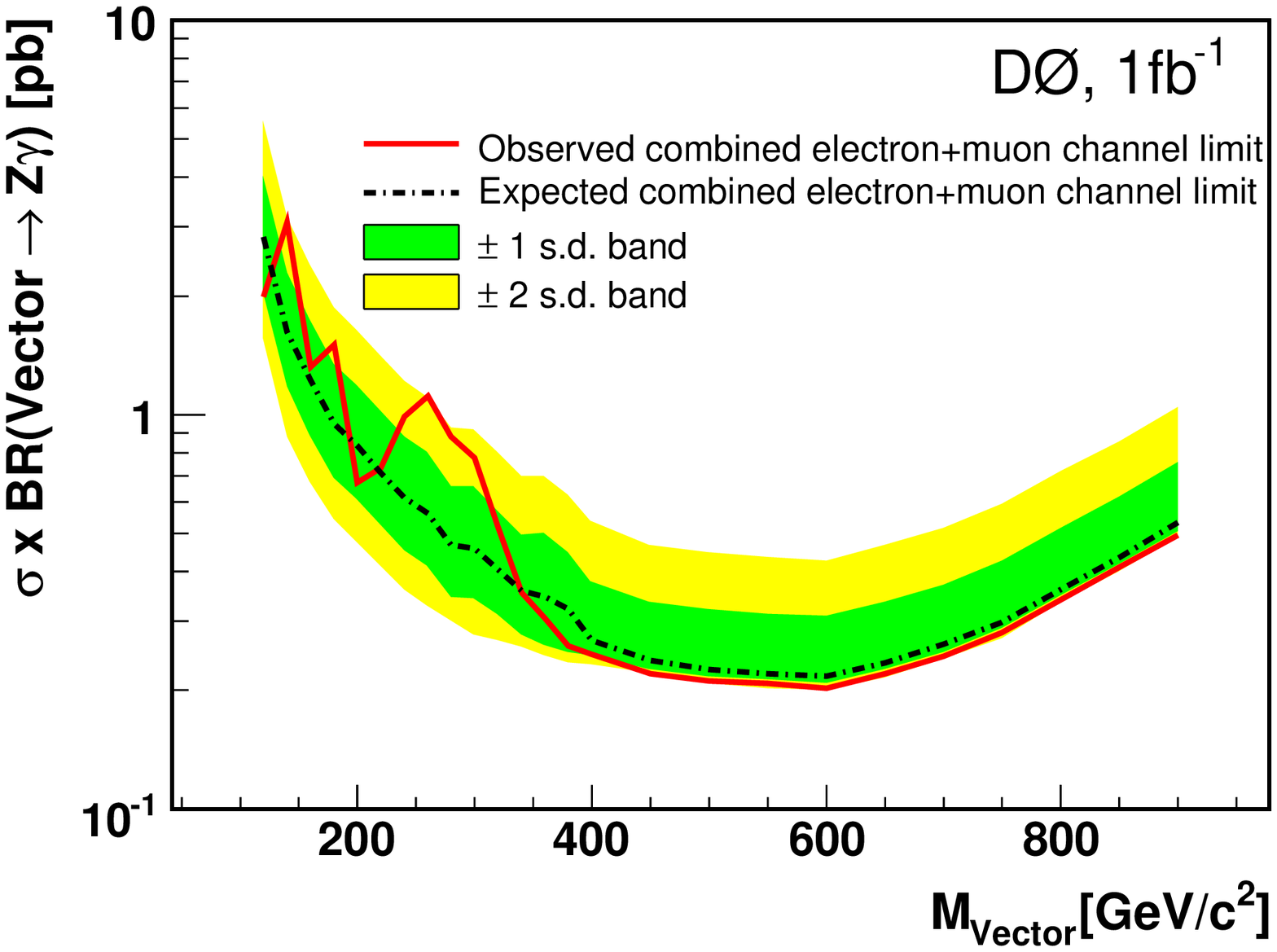}
\caption{The observed and expected $\sigma \times {\cal B}$ 95\% C.L. limits for a scalar (left) and vector (right)
particles decaying into Z$\gamma$ as a function of the resonance mass. The bands represent the 1 s.d. (dark) and 2 s.d. (light) uncertainties on the expected limit.
\label{fig:scalar_limit}}
\end{figure}

The combined $M_{\ell\ell\gamma}$ distribution from both channels is compared with the backgrounds and examined 
for any discrepancies (Fig.~\ref{fig:mllg_vector}). The $M_{\ell\ell\gamma}$ distributions from various MC 
signals of a vector resonance are also shown in Figure~\ref{fig:mllg_vector}. A thorough examination revealed no excess of the observed data over the theoretical expectations,
thus allowing us to set limits on the production of resonances decaying into $Z\gamma$. To set limits we explore the modified frequentist method~\cite{fisher} with Poisson log-likelihood ratio test statistic (LLR)~\cite{junk} with Gaussian uncertainties as the quantitative measure of the difference between the background-only 
hypothesis and the signal-plus-background one that predicts a $Z\gamma$ resonance. 
We derive 95\% C.L. upper limits on the cross section times branching fraction, that range from 0.19 (0.20) pb to 2.5 (3.1) pb for a scalar (vector) resonance. The resulting exclusion curves for $\sigma \times {\cal B}$ are shown in Figure~\ref{fig:scalar_limit} for the vector and scalar models. The branching fractions for the $Z$ boson to $ee$ or $\mu \mu$ are accounted for in these results.


\begin{acknowledgments}
Special thanks to my D0 Colleagues Y.~Maravin, V.~Cuplov, A.~Alton, R.~Averin for their help in preparing this poster. The author is grateful to the Editorial Board 016 chaired by G.~Landsberg, Electroweak working group chaired by T.~Bolton, H.~Schellman and J.~Zhu, and Diboson group chaired by A.~Lyon and A.~Askew for their help in reviewing the analysis. I would also like to thank S.~Mrenna and W.~Fisher for their help with the generator and limit setting tools.
\end{acknowledgments}



\begin{thebibliography}{99} 

\bibitem{Buescher:2005re} V.~Buescher and K.~Jakobs, Int.\ J.\ Mod.\ Phys.\ A {\bf 20}, 2523 (2005).
\bibitem{Djouadi} A.~Djouadi, J.~Kalinowski, and M.~Spira
  Comput. Phys. Commun. {\bf 108}, 56 (1998).
\bibitem{Kozlov:2005rj}  G.~A.~Kozlov, Phys.\ Rev.\ D {\bf 72}, 075015 (2005).
\bibitem{Ono:1983tf} S.~Ono, Acta Phys.\ Polon.\ B {\bf 15}, 201 (1984).
\bibitem{Cakir:2004nh}  O.~Cakir, R.~Ciftci, E.~Recepoglu and S.~Sultansoy,  Acta Phys.\ Polon.\ B {\bf 35}, 2103 (2004).
\bibitem{Hill} C.T.~Hill, E.H.~Simmons, Phys. Rept. {\bf 381}, 235 (2003) [Erratum-ibid. {\bf 390}, 553 (2004)].
\bibitem{us} D0 Collaboration, V.M. Abazov {\it et al.}, Phys. Lett. B {\bf 653}, 378 (2007).
\bibitem{Abazov:2006ez} D0 Collaboration, V.M. Abazov {\it et al.}, Phys. Lett. B {\bf 641}, 415 (2006); ``Erratum to
Search for Particles Decaying into a Z Boson and a Photon in $p\bar{p}$ Collisions at sqrt(s) = 1.96 TeV''.
\bibitem{plb_subm} D0 Collaboration, V.M. Abazov {\it et al.}, arXiv:0806.0611v1 [hep-ex].
\bibitem{pythia} T. Sj{\"o}strand {\it et al.}, Computer Physics Commun. {\bf 135}, 238 (2001).
\bibitem{madevent}F. Maltoni and T. Stelzer, JHEP {\bf 0302}, 027 (2003).
\bibitem{run2det} D0 Collaboration, V.M. Abazov {\it et al.}, Nucl. Instrum. Methods Phys. Res. A {\bf 565}, 463 (2006).
\bibitem{EM_isolation} {\it Isolation} $= \frac{E_{\rm tot}(R<0.4) - E_{\rm EM}(R<0.2)}{E_{\rm EM}(R<0.2)}$, where $E_{\rm EM}~(R<0.2)$ and $E_{\rm tot}(R<0.4)$ are the EM energies within a cone of radius $R = \sqrt{(\Delta \phi)^{2} + (\Delta \eta )^{2}} =$ 0.2 and 0.4, respectively. We require electrons (photons) to have {\it Isolation} $<$~0.20 (0.15).
\bibitem{emjets} EM-like jet is a jet with most of its energy carried by photons, and misidentified as an electron or a photon candidate.
\bibitem{baur} U. Baur and E. Berger, Phys. Rev. D {\bf 47}, 4889 (1993).
\bibitem{baur_NLO} U. Baur, T. Han and J. Ohnemus, Phys. Rev. D {\bf 57}, 2823 (1998).
\bibitem{fisher} W. Fisher, FERMILAB-TM-2386-E (2007).
\bibitem{junk} T. Junk, Nucl. Instrum. Methods Phys. Res. A {\bf 434}, 435 (1999).
\end{thebibliography}
\end{document}